%% file: main.tex
\documentclass[sigconf, authorversion,nonacm]{acmart}

\AtBeginDocument{%
  }

\begin{document}

\title{Large Language Models in Cybersecurity: State-of-the-Art}

\author{Farzad Nourmohammadzadeh Motlagh}
\affiliation{%
  \institution{Hasso-Plattner-Institute for Digital Engineering, University of Potsdam}
  \country{Germany}}
\email{farzad.motlagh@hpi.de}

\author{Mehrdad Hajizadeh}
\affiliation{%
  \institution{Technische Universitat Chemnitz}
  \country{Germany}}
\email{mehrdad.hajizadeh@etit.tu-chemnitz.de}

\author{Mehryar Majd}
\affiliation{%
  \institution{Hasso-Plattner-Institute for Digital Engineering, University of Potsdam}
  \country{Germany}}
\email{mehryar.majd@hpi.de}

\author{Pejman Najafi}
\affiliation{%
  \institution{Hasso-Plattner-Institute for Digital Engineering, University of Potsdam}
  \country{Germany}}
\email{pejman.najafi@hpi.de}

\author{Feng Cheng}
\affiliation{%
  \institution{Hasso-Plattner-Institute for Digital Engineering, University of Potsdam}
  \country{Germany}}
\email{feng.cheng@hpi.de}

\author{Christoph Meinel}
\affiliation{%
  \institution{Hasso-Plattner-Institute for Digital Engineering, University of Potsdam}
  \country{Germany}}
\email{christoph.meinel@hpi.de}

\renewcommand{\shortauthors}{Motlagh et al.}

\begin{abstract}
    \input{content/00_abstract}
\end{abstract}



\maketitle

\input{content/01_introduction}

\input{content/02_backgroun-related-work}

\input{content/03.01_main-defensive}

\input{content/03.02_main-offensive}

\input{content/04_future-work-and-conclusion}

\bibliographystyle{ACM-Reference-Format}
\bibliography{references}

\end{document}

%% file: content/00_abstract.tex
\textit{
The rise of Large Language Models (LLMs) has revolutionized our comprehension of intelligence bringing us closer to Artificial Intelligence. Since their introduction, researchers have actively explored the applications of LLMs across diverse fields, significantly elevating capabilities. Cybersecurity, traditionally resistant to data-driven solutions and slow to embrace machine learning, stands out as a domain.
This study examines the existing literature, providing a thorough characterization of both defensive and adversarial applications of LLMs within the realm of cybersecurity. Our review not only surveys and categorizes the current landscape but also identifies critical research gaps. By evaluating both offensive and defensive applications, we aim to provide a holistic understanding of the potential risks and opportunities associated with LLM-driven cybersecurity.
}

%% file: content/01_introduction.tex
\section{Introduction}
\label{intro}
The evolution of generative artificial intelligence, notably large language models (LLMs), has influenced most disciplines of science and technology that support content generation in diverse applications \cite{neupane2023impacts}. In education, LLMs support educators in various tasks such as assignment assessment \cite{hsiao2023developing}, question generation \cite{elkins2023useful}, providing feedback \cite{guo2023resist}, and essay grading \cite{yan2023practical}. In the entertainment industry, LLMs demonstrate competitive performance in generating music captions \cite{deng2023musilingo} as well as video game scripts \cite{latouche2023generating}. Automation is introduced into customer service \cite{pandya2023automating}, marketing \cite{gan2023making,yang2023against}, and supply chain management \cite{hendriksen2023ai,li2023large,kosasih2023review} through the integration of LLMs in business. Meanwhile, the utilization of LLMs in healthcare enables professionals by providing real-time clinical decision support \cite{rao2023evaluating,fawzi2023review}, medical education \cite{kuckelman2023assessing,song2023evaluating}, and prediction of disease progression \cite{shoham2023cpllm,rasmy2021med}. \par

With advancements in cyber threats, the cybersecurity domain can also be equipped with cutting-edge tools, assisting cybersecurity practitioners who continuously seek solutions to implement advanced policies or strengthen technological protections against the disclosure of confidential information, unauthorized access, and other forms of data modification \cite{kaur2023artificial}. Thanks to LLMs' capability in breaking down complex natural language patterns, security experts are now enabled to explore more attack vectors in various contexts associated with textual data \cite{yang2023harnessing}. \par
Functionalities of LLMs are increasingly being integrated into the cybersecurity posture, contributing to promising enhancements in cybersecurity defense applications \cite{li2023privacy}. Through analyzing vast amounts of text data, including security logs, these models can identify emerging vulnerabilities. Anomaly detection represents a key application of LLMs for identifying potential threats \cite{liu2023logprompt}. Furthermore, LLMs mitigate potential risks by offering automated vulnerability fixes, aiming to improve organizations' security posture \cite{pearce2023examining}. \par

However, with the continuous advancements of LLMs in cyber defense, it is crucial to acknowledge that these language models can also be leveraged by malicious actors. For example, LLMs can be misused by attackers to execute malware in target companies \cite{botacin2023gpthreats}, engage in defense evasion \cite{chatzoglou2023bypassing}, and gain access to credentials \cite{rando2023passgpt}. The potential to generate complex and personalized phishing messages further highlights the misuse of LLMs for deceiving people in an organization, paving the way for unauthorized access to companies' sensitive information \cite{saha2023generating}. To further elaborate, WormGPT \cite{falade2023decoding} is an AI-powered tool designed for cybercriminals to automate the generation of personalized phishing emails. Although it may sound somewhat similar to ChatGPT, WormGPT is not a friendly neighborhood AI; instead, its purpose is to produce malicious content. Furthermore, FraudGPT \cite{Dutta_2023} enabled attacker to create content to convince users to click on a particular generated link.\par

The dual nature of LLMs has transformed the cybersecurity realm by offering new challenges and opportunities. Developing robust defensive strategies to foresee attacks and address concerns related to the utilization of LLMs motivated us to formulate a taxonomy of strategies appearing in the field of cybersecurity. To define our contributions more precisely, this paper addresses:
\begin{itemize}

\item The intersection of LLMs' offensive approaches as a newly introduced dimension to cybersecurity is framed in this study in line with the Mitre attack framework \cite{MITRE2022}.
\item Exploring LLM-empowered defensive strategies in dealing with potential threats and malware based on the NIST cybersecurity framework \cite{cybersecurity2014framework}.
\item Understanding the major functionalities of LLMs in current research trends alongside potential applications in the cybersecurity landscape.

\end{itemize}
The rest of paper is organized as follows: In Section \ref{background}, we provide an overview of LLMs . Moving forward to Section \ref{defense}, we explore cyber threat defenses leveregred by LLMs where Section \ref{attack} outlines  sophisticated attacks designed by LLMs. Finally, Section \ref{conclusion} concludes the challenges posed by LLMs in the context of cybersecurity.

%% file: content/02_backgroun-related-work.tex
\section{Background}
\label{background}
LLMs are neural networks that learn from textual data to process various language-related tasks \cite{naveed2023comprehensive}. From Eliza as a pattern recognition chatbot in the 1960s \cite{weizenbaum1966eliza}, over the years several advancements pushed Natural Language Processing (NLP) forward, such as long short-term memories to handle a wide range of data \cite{hochreiter1997long}, Stanford CoreNLP suite \cite{manning2014stanford} providing a collection of algorithms to perform intricate NLP tasks and continued with transformer architecture \cite{vaswani2017attention}. \par
 
A breakthrough in Transformer-based models surged the field of NLP and led to the development of numerous kinds of effective LLMs. T5 \cite{raffel2020exploring} applied language modeling in pre-trained LLMs, where spans are altered with a single mask. GPT-3 enhanced the performance of LLMs with size by increasing model parameters to 175B. PaLM-2 is trained on high-quality datasets \cite{anil2023palm} with an objective of cutting the cost of training and inference \cite{naveed2023comprehensive}. Llama, a set of decoder-only models aimed at minimizing the amount of activations in the backward step \cite{naveed2023comprehensive,touvron2023llama}. Xuan Yuan 2.0, a Chinese financial chat model  \cite{naveed2023comprehensive,zhang2023xuanyuan}, AlexaTM \cite{soltan2022alexatm}, PaLM-2 \cite{anil2023palm}, as well as GLM-130B \cite{zeng2022glm} are a few instances of general purpose pre-trained LLMs. While pre-trained models offer an essential understanding of languages, as AI advances, fine-tuning LLMs boost business functions and satisfaction by fulfilling industry-specific criteria \cite{zhang2023balancing}. A general-purpose LLaMA-GPT-4 \cite{peng2023instruction}, Goat \cite{liu2023goat} for handling complicated arithmetic queries, HuaTuo \cite{wang2023huatuo} a medical knowledge model, Evol-Instruct \cite{xu2023wizardlm} offering complicated prompts, and LLaMA 2-Chat fine-tuned using rejection sampling \cite{touvron2023llama} are exemplary instruction-tuning LLMs.
 Running in higher costs, extensive hardware requirements, cost of slow training on various tasks, limited LLMs utilization \cite{naveed2023comprehensive}. Retrieving support evidence from an external in-domain knowledge base \cite{zhang2023reformulating},  parameter tuning and knowledge distillation are among the techniques extensively researched for effective LLM deployment \cite{naveed2023comprehensive}. \par
 
 Recently, the scientific literature has experienced a significant growth in the number of articles related to LLMs, principally driven by their proven efficacy across a wide range of functions. As a result, throughout various surveys, researchers attempted to categorize these advancements in LLM architecture \cite{naveed2023comprehensive,zhao2023survey,zhou2023comprehensive,huang2022towards}. Though previous studies have investigated literature reviews to highlight the safety aspects of LLMs \cite{iturbe2023artificial,addington2023chatgpt,kucharavy2023fundamentals,ishihara2023training}, the present study focuses primarily on the application of LLMs in the context of cyberdefense as well as cyberattack.\par

%% file: content/03.01_main-defensive.tex
\section{Defensive Applications of LLMs}
\label{defense}
In the field of cybersecurity, the National Institute of Standards and Technology (NIST) provides a comprehensive structure to enhance organizations’ cybersecurity status, as detailed in the NIST cybersecurity framework  \cite{cybersecurity2014framework}. According to its effectiveness and popularity in cyberdefense, we classify the diverse array of LLM-centered approaches that contributed in cyberdefense through the lens of NIST framework to better understand the impact of LLMs in cyberdefense. The framework consists of a structured approach to identify, protect, detect, respond to, and recover from cybersecurity threats and incidents.

\subsection{Identify}

The process of developing an organizational understanding to manage cybersecurity risk concerning systems, assets, data, and capabilities is referred to the \textit{Identify} function in the context of the NIST framework \cite{cybersecurity2014framework}. Identifying potential risks is a crucial phase in risk management, and LLMs aim to fulfill a transformational role in forming risk management in businesses. Johnson \cite{Medium} presents invaluable insights for policymakers on the applicability of LLMs to risk management. According to the author, LLMs go through business headlines, social media posts, economic indicators, legal documentation, and other key sources, emphasizing risk elements to deliver more accurate and predictive risk assessments that a human analyst might overlook. Lima et al. \cite{de2023learning} develop a risk matrix from application reviews using LLMs. Through user feedback, they proposed an automatic prompt extraction technique. These prompts were passed into LLMs, which classified the risks into five classes ranging from negligible to critical for further investigation. Naleszkiewicz \cite{Naleszkiewicz_2023} discusses LLM applications allowing companies to overcome traditional enterprise risk management challenges, such as operational and compliance risks. LLMs evaluate unstructured siloed data across various departments, acting as a bridge to provide an in-depth understanding of an organization's risk profile. Furthermore, LLMs boost risk modeling by generating expert opinions based on prior patterns, risk mitigation by generating contingency plans, and risk reporting by providing customized risk assessments.\par

\subsection{Protect}
Implementing safeguards to guarantee the delivery of essential services is reflected in \textit{protect} function \cite{cybersecurity2014framework}. It involves various mechanisms such as maintaining a proactive security posture or prioritizing cybersecurity awareness and training to empower the organization's workforce. In the current digital environment, proactive protection technologies are essential since they enable companies to anticipate and prevent troubles before they arise. For example, proactive technologies empower enterprises to minimize the likelihood of coming across inappropriate content,  and thus reduce the possibility of experiencing ethical or legal challenges \cite{sun2023cyber}. Voros et al. \cite{voros2023web} harnessed the power of LLMs to enhance web content filtration. They have improved the accuracy of web content categorization by scanning of large amount of URLs. Another research accomplished by Yu et al. \cite{yu2023honey} investigates GPT-3's capacity to produce honeywords to trap the attackers if they are using deceptive generated passwords. First, they extract the components of the original password using a password-specific segmentation algorithm. These segments are then fed into GPT-3 as a prompt to generate a collection of passwords similar to the input password. A crucial element in this model's efficacy is the maintenance of strong password components called chunks given to the LLM \cite{sannihith2023enhancing}.\par
LLMs can play a valuable role in strengthening cybersecurity awareness and training within the protect function of the NIST framework. Tann et al. \cite{tann2023using} apply LLMs to tackle professional certification topics and perform Capture The Flag (CTF) tasks to improve participants’ cybersecurity education. LLMs have significance by enabling attendees to explore CTF test settings, providing explanations to concerns connected to professional certification, and highlighting the need to model cybersecurity breach scenarios in CTF sessions to support the development of more comprehensive skills. However, LLMs face limitations when it comes to responding to conceptual queries. Furthermore, LLMs can improve team collaboration by offering security question solutions that are suitable for inexperienced as well as experts. For instance, LLMs greatly increase the efficacy of penetration test teams by making it easier for team members to pass on information by offering more in-depth assessments and generating appropriate explanations to be on the same page about the detected risks. Moreover, LLMs serve as a connection between experts and publicly accessible web resources, in particular assisting specialists in remaining up to date on the most recent security concerns that are critical to their company \cite{dutta2018utilizing}. \par


Automated vulnerability fixing with LLMs diminishes the risk of cyberattacks. A three-step process is described by Charalambous et al. \cite{charalambous2023new} for addressing automotive vulnerability issues. Bounded Model Checking (BMC) is the first step in the process. It evaluates the user-provided source code to a property specification. The original code and the appropriate counterexample are provided to the LLM module by the BMC engine in the scenario that this phase's verification is unsuccessful and a security property violation is detected. Secondly, customized queries are sent to the LLM engine to produce a corrected version of the code. Lastly, the BMC module re-evaluates the code that the LLM module changed to formally determine whether the updated version matches the original security and safety requirements.\par

Automating flaw mitigation can be facilitated by LLMs if the defect is well-defined and the prompt provides additional information. While these models were fully effective in fixing simulated vulnerabilities, real-world scenarios presented challenges for their performance. The primary challenges stem from the numerous methods that information is presented, the complexities of prompt processing and code development in LLMs, and the significance of prompt phrasing, which can result in notable variations in the code required to be generated \cite{pearce2023examining}. Furthermore, Sandoval et al. \cite{sandoval2023lost} performed an examination of potentially insecure code suggestions during the process of code development. Within a particular programming context that the authors had defined, they tested scenarios with and without AI support. Their findings indicate that users assisted by AI develop security flaws at a rate lower than ten percent, suggesting that using LLMs in their security-oriented research does not present major new security risks.\par



\subsection{Detect}
The NIST framework's \textit{Detect} function serves to identify cybersecurity events as they arise \cite{cybersecurity2014framework}. Exploring anomaly detection in system logs is a crucial step toward developing effective detection methods through the use of LLMs. Recurrent Neural Network Language Models are used by Tuor et al. \cite{tuor2018recurrent} to present an unsupervised, online anomaly detection method for computer security log analysis. This approach simplifies the usual effort-intensive feature engineering stage, making it fast to implement, and is independent of the tools used for system configuration and monitoring. The authors have demonstrated the efficacy of their approach by utilizing the Los Alamos National Laboratory Cyber Security Dataset \cite{kent2016cyber}. Their findings indicate that the approach can be handled in real-time, generating and organizing log-line-level anomaly scores while taking into account inter-log-line context. The authors \cite{tuor2018recurrent} considered metrics including Average Percentile (AP) and Area under the Receiver Operator Characteristic Curve (AUC) to show how the false-positive rate dropped without significantly affecting the ability to detect unusual behavior \cite{kent2016cyber}. \par
GPT-2 is used by VulDetect\cite{omar2023vuldetect}, a transformer-based vulnerability detection framework, to detect anomalies in system logs. Using a dataset containing both vulnerable and non-vulnerable code, the model is fine-tuned to detect anomalies that represent regular behavior. Malicious behavior is defined as any unexpected or unlikely outcome that the model possibly generated.
Two benchmark datasets, SARD \cite{zhou2022vulnerability} and SeVC \cite{shoeybi2019megatron}, were utilized by the authors to assess VulDetect's performance. The outcomes showed that VulDetect has a low false positive rate and is efficient in real-time vulnerability detection.
Moreover, the integration of LLMs into penetration testing practices has the potential to revolutionize the world of threat detection. Threat detection could undergo a revolution if LLMs are incorporated into penetration testing procedures. Happe et al.'s investigation \cite{happe2023getting} focused on using LLMs to improve penetration testing. In line with their classification, LLMs provide advancement in two aspects of penetration testing: high-level and low-level operations. High-level assignments include conceptual investigation and strategic planning, such as finding out about emerging active directory attacks. On the other hand, tasks at a lower level incorporate consideration of practical activities involving system exploitation and vulnerability analysis. This entails looking for specific attack vectors for a particular system.\par
A further investigation by Deng et al. \cite{deng2023pentestgpt} introduces PENTESTGPT, an automated penetration testing system driven by LLMs. Complex tasks such as question answering, summarization, and reasoning are readily handled with PENTESTGPT. Addressing context loss concerns and simulating human behavior in penetration testing are the objectives. Three self-interacting modules jointly form PENTESTGPT including reasoning, generation, and parsing. These modules collaborate to tackle penetration testing problems by using a divide-and-conquer approach. Specific subtasks are allocated to each module, which interact to effectively handle and compile the data generated during testing.\par
Ranade et al. \cite{ranade2021cybert} improve the processing of threats, attacks, and vulnerabilities which is challenging due to the high volume of data, and the dynamic nature of evolving attack techniques. The primary objective of their research is an enhanced version of a BERT model, which aims to effectively perform several cybersecurity-related operations. Using Masked Language Modeling (MLM), the model was trained using unstructured and semi-structured open-source Cyber Threat Intelligence (CTI) data. Its evaluation encompassed diverse downstream tasks with potential applications in Security Operations Centers (SOCs). They additionally offer real-world examples of how to apply CyBERT to cybersecurity problems. Several subsequent works have furthered the advancements of this research in terms of both training efficiency and accuracy such as SecureBERT \cite{aghaei2022securebert}, CySecBERT \cite{bayer2022cysecbert}, and ClaimsBERT \cite{ameri2022accuracy}. In this regard, Bayer et al. \cite{bayer2022cysecbert} presented a word embedding model based on BERT and collected a dataset from multiple sources. This adaptation makes the model capable of coping with a wide range of cybersecurity tasks, namely malware detection, alert aggregation, and phishing website detection.\par
The LLMs can also facilitate auditory tasks to detect vulnerabilities among the smart contracts. David et al. \cite{david2023you} utilized LLMs to target vulnerabilities in the smart contracts and DeFi protocol layers. Their study detects 52 compromised DeFi protocols, as input data for the language model context, evaluating the impact of model temperature and context length on the language model's efficacy in smart contract auditing.
The results indicated that incorporating LLMs into the audit workflow substantially boost the effectiveness and accuracy of analyzing an array of feasible attacks. On the other hand, Chen et al. \cite{chen2023chatgpt} trained LLM on a dataset of 10,000 smart contracts and evaluated how well it detected nine different vulnerabilities. According to the authors' findings, LLMs frequently deliver false positive results when detecting smart contract vulnerabilities. This might be connected with interference from incomplete codes or LLMs' incapacity to understand code segments.\par
 An LLM can be used to build a scenario comparable to an attacker's strategy for gaining access to an organization's property by exploiting a vulnerability. Garvey et al. \cite{garvey2023can} study the viability of using Generative-AI to improve the development of Red Team scenarios in organizations. The authors \cite{garvey2023can} propose employing LLMs to construct narratives based on prompts or questions as input. Subsequently, subject-matter specialists provide remarks, including modifying narratives, adding new elements, or integrating multiple items to develop more complex scenarios. The objective is to guarantee that the generated scenarios are plausible and adhere to the provided framework. They found that including elements inspired by fiction into LLMs improves creativity and imagination in the scenario development process.\par
Koide et al. \cite{koide2023detecting} present a strategy for detecting phishing websites using LLMs. Their approach entails using a web crawler to retrieve data from websites and creating prompts for LLMs. Social engineering strategies are then identified by evaluating the context of entire web pages and URLs. The prompts rely on the Chain of Thought (CoT) prompting technique, which enables LLMs to elaborate on their reasoning. In addition, the study recommends an HTML simplification approach to improve efficiency. This entails lowering the token count by simplifying HTML text and removing HTML elements that lack text within tags, such as style, script, and comment tags. This operation is repeated until the token count reaches a certain threshold, thus boosting overall efficiency.\par

Sakaoglu introduced KARTAL\cite{sakaoglu2023kartal}, a fine-tuned Language Model for detecting vulnerabilities in web applications. A detector component in the KARTAL system is controlled by the prompts from the prompter component. These prompts are generated based on input gathered by the fuzzer component, which monitors application activity. The LLM detects logical vulnerabilities in web applications, specifically broken access control rules, by analyzing these prompts. This technique allows KARTAL to dynamically alter the definitions of broken access, allowing it to adapt to a variety of scenarios. This adaptability distinguishes it from less intelligent vulnerability scanners, allowing KARTAL to be more effective in its detection capabilities.

LLMs demonstrate their capacity to be an effective method across a wide range of vulnerability identification tasks. CyBERT \cite{ameri2021cybert} unveils a classifier for detecting cybersecurity feature claims. The method incorporates fine-tuning a pre-trained BERT language model to recognize cybersecurity claims throughout complex sequences observed in industrial control systems (ICS) device documentation. This is accomplished by aggregating reports for each feature from every source linked with an individual device, effectively determining in-conflict feature claims. The extraction of sequences from ICS-related documents is the initial stage in the procedure as these sequences are classified into broad claims, device claims, or cybersecurity claims. Then, the identified sequences are used to train CyBERT so it can classify new sequences.\par

SecurityLLM, a system developed for precise threat detection and data privacy, is presented by Ferrag et al. \cite{ferrag2023revolutionizing}. SecurityLLM utilizes Fixed-Length Language Encoding (FLLE) as a privacy-preserving encoding method, in conjunction with the Byte-level Byte-Pair Encoder (BBPE) Tokenizer forming text traffic data. The SecurityLLM framework is composed of two primary components: SecurityBERT, which detects cyber threats, and FalconLLM, which responds to and recovers from incidents. The method, which was trained on an IoT cybersecurity dataset, displays significant accuracy in identifying fourteen various types of cyber threats.\par

SecureFalcon \cite{ferrag2023securefalcon} is an LLM-based cybersecurity reasoning system targeted to detect software flaws. The method involves fine-tuning FalconLLM with the use of a FormAI dataset including C code instances. SecureFalcon \cite{ferrag2023securefalcon} uses binary classification to distinguish between vulnerable and non-vulnerable patterns and then validates corrected code using Bounded Model Checking. However, the study's adaptability is limited due to the FormAI dataset's exclusive focus on C codes.\par
\subsection{Respond}
The \textit{Respond} function involves the formulation of actions to address the detected incident \cite{cybersecurity2014framework}. The convergence of LLMs and honeypot paradigms enhances the capability to respond to malware threats. In exploring this synergy, McKee et al. \cite{mckee2023chatbots} research the feasibility of using LLMs to improve cybersecurity in a honeypot setup. The researchers \cite{mckee2023chatbots} demonstrate how these chatbots can create a responsive honeypot interface capable of responding to illicit activities. This method gives security professionals more time to respond to an ongoing cyber attack. Ten tasks connected with the development of honeypots are divided into three primary categories by the authors \cite{mckee2023chatbots}: networks, operating systems, and applications. Their results indicate that the LLM-based honeypot interfaces are able to maintain the attacker's interest over the course of several inquiries. In another study, Sladic et al. \cite{sladic2023llm} present an LLM-based technique for developing software honeypots. The devised honeypot named shelLM is designed to evaluate the credibility of the model through the use of security experts in an experiment.  The specialists collaborated with ShelLM to assess how it responded to the commands of an attacker. ShelLM's ability to retain consistency over several sessions is a significant feature; the content of each terminal session is kept and used as a prompt for following sessions. This makes sure that regardless of when a session comes to an end interactions can carry on without interruption. Cambiaso et al. \cite{cambiaso2023scamming} deliver a method for generating email messages to identified attackers in order to engage them and squander their resources. LLMs provide realistic responses based on human behavior, making scams less profitable. However, such automated responses need a significant amount of storage and computational power. \par

\begin{table*}[t]
\caption {
Classified publications concerning the \textit{defensive}  applications of LLMs.
}
\label{TableDefense}
\centering
\footnotesize
\resizebox{0.95\textwidth}{!}{
\begin{tabular}{lcclc} 
\hline

\textbf{Paper} &
\textbf{Year} &
\textbf{NIST Framework} &
\textbf{Application} &
\textbf{Model(s)} \\
\hline
\cite{kereopa2023building}   & 
2023 &
Identify &
\begin{tabular}[c]{@{}l@{}}LLMs enhance cybersecurity policies.\end{tabular}&
ChatGPT \\
\hline


\cite{he2023large}   &
2023 &
Protect &
\begin{tabular}[c]{@{}l@{}}Using LLMs for secure code development without compromising functionality.\end{tabular}&
\begin{tabular}[c]{@{}c@{}}SVEN (GPT-2),\\  (CodeGen) LM \end{tabular}\\
\hline

\cite{tann2023using}   &
2023 &
Protect &
\begin{tabular}[c]{@{}l@{}}LLMs solve Capture The Flag challenges to enhance employees' \\awareness and knowledge.\end{tabular}&
\begin{tabular}[c]{@{}c@{}}code-cushman-001,\\ code-davinci-001,code-davinci-002,\\ 1-jumbo, j1-large, polycoder, gpt2-csrc \end{tabular}\\
\hline

\cite{pearce2023examining}   &
2023 &
Protect &
\begin{tabular}[c]{@{}l@{}}LLMs investigate software vulnerabilities.\end{tabular}&
\begin{tabular}[c]{@{}c@{}}GPT-3.5 Turbo,\\ Gemini,\\ Microsoft Bing \end{tabular}\\
\hline

\cite{charalambous2023new}   &
2023 &
Protect &
\begin{tabular}[c]{@{}l@{}}LLMs investigate software vulnerabilities.\end{tabular}&
GPT-3.5 Turbo \\
\hline

\cite{yu2023honey}   &
2023 &
Protect &
\begin{tabular}[c]{@{}l@{}}Generating honeywords using LLMs.\end{tabular}&
GPT-3 \\
\hline

\cite{dutta2018utilizing}   & 
2018 &
Protect &
\begin{tabular}[c]{@{}l@{}}Chatbots assist security experts in identifying open ports.\end{tabular}&
Rule-based \\
\hline

\cite{voros2023web}   &
2023 &
Protect &
\begin{tabular}[c]{@{}l@{}}LLM-based URL categorization for website classification.\end{tabular}&
\begin{tabular}[c]{@{}c@{}}eXpose (Conv),\\ BERTiny, URLTran (BERT)\\ T5 Large, GPT3 Babbage \end{tabular}\\
\hline

\cite{sandoval2023lost}   &
2023 &
Protect &
\begin{tabular}[c]{@{}l@{}}LLMs investigate code vulnerabilities.\end{tabular}&
GPT-3 \\
\hline


\cite{tuor2018recurrent}   &
2018 &
Detect &
\begin{tabular}[c]{@{}l@{}}Detecting anomalous behavior in network logs with LLMs.\end{tabular}&
RNN \\
\hline

\cite{omar2023vuldetect}   &
2023 &
Detect &
Detection of vulnerabilities in software code. &
GPT-2 \\
\hline

\cite{gao2023advance}  &
2023 &
Detect &
\begin{tabular}[c]{@{}l@{}}SecureBERT for anomaly detection.\end{tabular}&
\begin{tabular}[c]{@{}c@{}}CyBERT,\\ SecureBERT (RoBERTa) \end{tabular}\\
\hline

\cite{ranade2021cybert}   & 
2021 &
Detect &
\begin{tabular}[c]{@{}l@{}}CyBERT, a domain-specific BERT model\\ to recognize specialized cybersecurity entities.\end{tabular}&
BERT-based Natural Language Filter  \\ 
\hline

\cite{happe2023getting}   &
2023 &
Detect &
\begin{tabular}[c]{@{}l@{}}Penetration testing with LLMs.\end{tabular}&
GPT-3.5  \\ 
\hline

\cite{ameri2021cybert}   &
2021 &
Detect &
\begin{tabular}[c]{@{}l@{}}CyBERT, a cybersecurity feature claims classifier.\end{tabular}&
CyBERT, GPT-2  \\ 
\hline

\cite{bayer2022cysecbert}   &
2022 &
Detect &
\begin{tabular}[c]{@{}l@{}}CySecBERT for malware detection and alert aggregation.\end{tabular}&
CySecBERT  \\ 
\hline

\cite{bayer2022cysecbert}   &
2022 &
Detect &
\begin{tabular}[c]{@{}l@{}}SecureBERT for processing and understandin cybersecurity text,\\ specifically Cyber Threat Intelligence (CTI).\end{tabular}&
SecureBERT  \\ 
\hline

\cite{ferrag2023securefalcon}   &
2023 &
Detect &
Detection of vulnerabilities in software code. &
SecureFalcon (FalconLLM)  \\ 
\hline

\cite{sladic2023llm}   &
2023 &
Respond &
Creating honeypots related to continuously monitoring and detecting threats. &
GPT-3.5 Turbo (shelLM)  \\ 
\hline

\cite{mckee2023chatbots}   &
2023 &
Respond &
\begin{tabular}[c]{@{}l@{}}LLM as a honeypot interface against command-line attacks.\end{tabular}&
GPT-3.5  \\ 
\hline

\cite{garvey2023can}   &
2023 &
Detect &
\begin{tabular}[c]{@{}l@{}}investigates LLMs acting as red teamers in cybersecurity.\end{tabular}&
GPT-4 \& Bard  \\ 
\hline

\cite{koide2023detecting}   &
2023 &
Detect &
\begin{tabular}[c]{@{}l@{}}LLM for detecting phishing sites leverages a web crawler\\ to gather information and generate prompts.\end{tabular}&
GPT-3.5 \& GPT-4  \\ 
\hline

\cite{sakaoglu2023kartal}   &
2023 &
Detect &
\begin{tabular}[c]{@{}l@{}}KARTAL, a web application vulnerability detection.\end{tabular}&
GPT-3.5 Turbo \\ 
\hline

\cite{david2023you}   &
2023 &
Detect &
\begin{tabular}[c]{@{}l@{}}LLMs to perform security audits on smart contracts.\end{tabular}&
\begin{tabular}[c]{@{}c@{}}GPT-4 (GPT-4-32k),\\  Claude-v1.3-100k \end{tabular}\\
\hline

\cite{deng2023pentestgpt}   &
2023 &
Detect &
\begin{tabular}[c]{@{}l@{}}LLM-empowered automatic penetration testing tool.\end{tabular}&
\begin{tabular}[c]{@{}c@{}}PentestGPT\\ (GPT-3.5 \& GPT-4) \end{tabular}\\
\hline

\cite{chen2023chatgpt}   &
2023 &
Detect &
\begin{tabular}[c]{@{}l@{}}LLMs to perform security audits on smart contracts.\end{tabular}&
\begin{tabular}[c]{@{}c@{}}GPT-3.5 Turbo \& GPT-4 \end{tabular}\\
\hline

\cite{cambiaso2023scamming}  &
2023 &
Respond &
\begin{tabular}[c]{@{}l@{}}Replying to the scam emails using LLM.\end{tabular}&
GPT-3 \\ 
\hline

\end{tabular}

} 
\end{table*}

\begin{figure}[t]
\includegraphics[width=1\columnwidth]{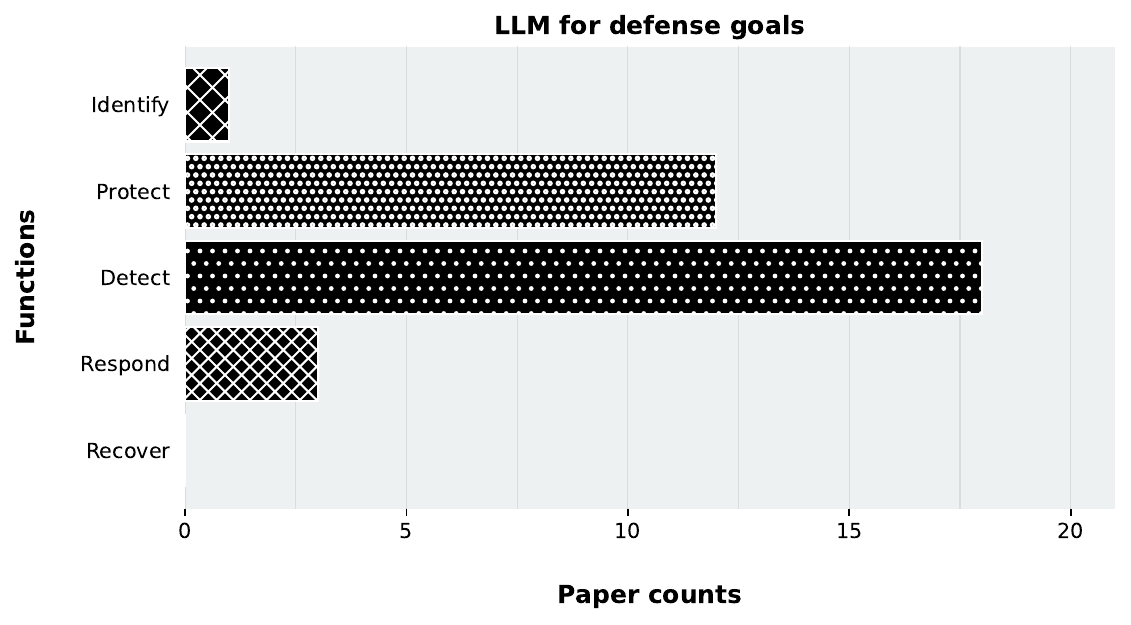}
\caption{The present bar chart illustrates the distribution of studies mapped to each of the five elements of the NIST Cybersecurity Framework. Collected statistics indicate that the vast amount of studies are related to Protect and Detect functions emphasizing research gaps related to Identify, Respond and particularly Recover functions over collected publications.}\label{fig2}
\end{figure}

We provide a set of insights based on existing work in Table \ref{TableDefense}. The present pattern of published papers on the use of LLMs for cyber defense indicates that most studies are focused on the detection and protection roles of LLMs aligning with the NIST framework. However, a research gap, as shown in Figure \ref{fig2}, becomes evident in post-attack scenarios. Given the critical roles recovery and attack response play in the cybersecurity lifecycle, it is essential that further studies be centered around the development of innovative LLM-related solutions to maximize their potential in productive post-attack scenarios.\par

%% file: content/03.02_main-offensive.tex
\section{Addversarial Application of LLMs}
\label{attack}
Applications of LLMs in cybersecurity extend beyond techniques for defense.  In our exploration, we review LLMs' capacity to come up with sophisticated attacks. To this end, our approach involves with analyzing these approaches through the MITRE attack framework, which outlines various attacker tactics.
\subsection{Reconnaissance}
During a reconnaissance attack, adversaries actively or passively collect information about their target organization in order to identify upcoming operations \cite{xiong2022cyber}. Hazell \cite{hazell2023large} provides an illustration of how LLMs assist during the reconnaissance stage by automating the data collection and analysis of potential victims. As a result, LLMs develop Python scripts to scrape websites that hold the desired information about users. Comparably, Salewski et al. \cite{salewski2023context} enabled the LLMs to assume various roles by introducing the prompt with "If you were a {persona}", in which the target individual is substituted for the persona. \par
\subsection{Initial Access}
The initial access tactic includes the procedures adopted by attackers to obtain access as a foothold to a company's infrastructure \cite{xiong2022cyber}. Roy et al. \cite{saha2023generating} highlight the role of LLMs in delivering malicious scripts where the attack structure is divided into four steps. In this regard, design objects are used to create concepts that are influenced by specific organizations, while credential-stealing objects are used to establish objects that require credentials, including login buttons or input fields. Credential Transfer objects are used to create functions that can provide the attacker with the credentials submitted on phishing websites. Lastly, the exploit generation object serves to implement a functionality based on the evasive exploit. The authors \cite{saha2023generating} conduct a number of attacks, including text encoding, clickjacking, polymorphic URL, and QR code-based multi-stage attacks, to show how LLMs have the potential to be leveraged to generate a variety of phishing attack forms. \par

According to Hazell et al. \cite{hazell2023large}, LLMs are able to assist during the reconnaissance stage of a spear phishing attack, a process when attackers get sensitive information about their targets in order to develop compelling messages.  According to John et al. \cite{john1802generative}, ML-based techniques group people according to their value and level of participation, and then utilize the timeliness of the target users to provide content and a phishing URL. Since people can adopt different personas in daily life and choose a variety of terms for a variety of circumstances, Kreps et al. \cite{kreps2022all} discuss how GPT2 can manipulate target users' beliefs by generating stories, while Salewski et al. \cite{salewski2023context} investigate the role of LLMs on various personas and adapt their language accordingly a process known as in-context impersonation. Based on LLMs ability to impersonate certain personalities, Salewski et al. \cite{salewski2023context} concluded that LLMs can be applied to develop more effective phishing messages or social engineering attacks. With a dataset of phishing emails, Karanjai \cite{karanjai2022targeted} investigates the effectiveness of generating convincing phishing emails with GPT2, GPT-3, and LSTM while taking into account the removal of HTML elements, URLs, and email addresses as well as tokenizing the text into words. \par

PassGPT, an LLM-based approach to password generation and modeling for password estimation, is presented by Rando et al. \cite{rando2023passgpt}. PassGPT presents the idea of guided password generation, enabling the generation of passwords that adhere to established standards. Moreover, PassGPT, trained on password leaks, models each token independently, a character-by-character search space exploration in which generated passwords are sampled according to random restrictions. \par

The application of LLMs, particularly ChatGPT and AutoGPT, in malware generation is covered by Pa Pa et al. \cite{pa2023attacker}. To determine if Auto-GPT minimizes the obstacle to malware generation, the authors \cite{pa2023attacker} investigated Auto-GPT running locally and tested it in the following manners: initially, by providing broad prompts like "write a malware X," and next, by giving more specific malware and attack tool functionalities. Finally, additional tests have been explored to discover whether Anti-Virus (AV), Endpoint Detection and Response (EDR), and VirusTotal (VT) detect the generated malware.\par

\subsection{Execution}
Procedures resulting in adversary-controlled executable operating on a local or remote system are referred to as execution \cite{xiong2022cyber}. Using code generation tools to develop malware is one of the strategies employed by adversaries. The feasibility of employing large textual models to automatically generate malware along with the model's constraints when generating actual malware samples is studied by Botacin \cite{botacin2023gpthreats}. According to their findings, certain malware versions were recognized by all antivirus engines while others were not detected by any of the engines due to the use of LLMs to modify all or part of the malware's building blocks. The prompt engineering essential to develop malware that hides a PowerShell and schedules its daily execution at a given time was brought to light by Charan et al. \cite{charan2023text}. In addition to copying the CMD file to a designated directory and getting the scheduled task information as a successful malware verification, the script adds a registry value that will be run at system startup. The LLM-based malware is assessed by Pa pa et al. \cite{pa2023attacker}. The authors \cite{pa2023attacker} reported that a number of the commercially available antivirus applications and Endpoint Detection and Response (EDR) solutions failed to detect the LLM-generated executables since some LLM-generated functions can establish connections toward attackers through the victim's machine \cite{beckerich2023ratgpt}.\par
\begin{figure}[ht]
\includegraphics[width=1\columnwidth]{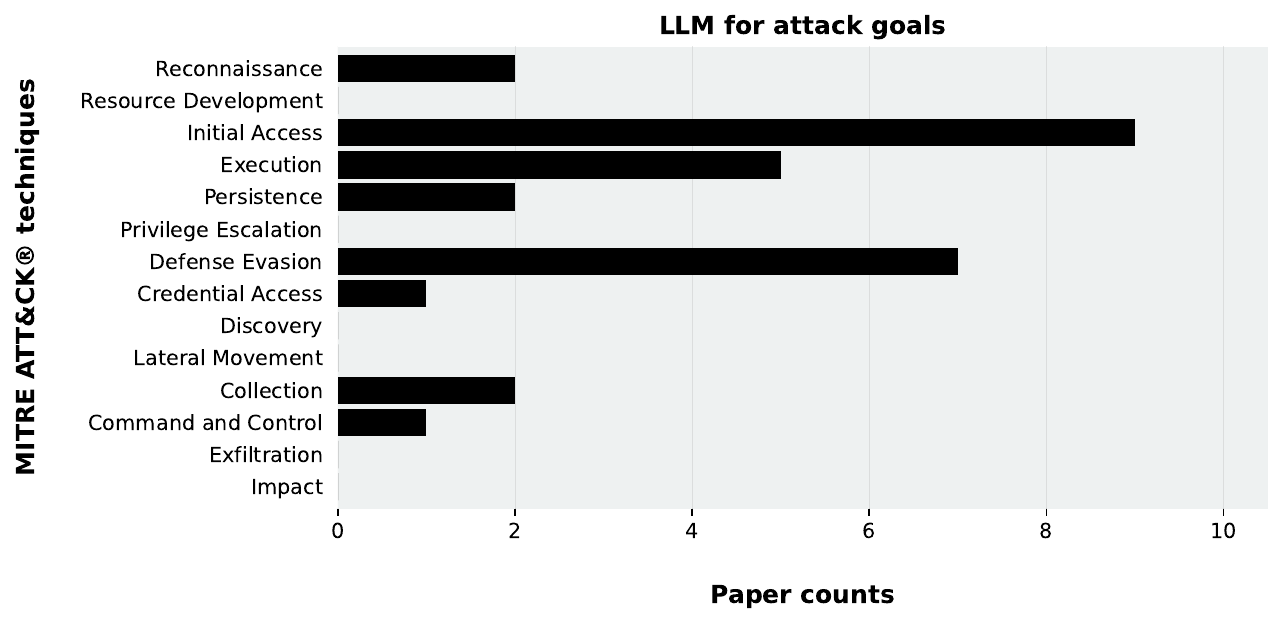}
\caption{Concentration of recently published papers on attack approaches using LLM}\label{fig3}
\end{figure}


\begin{table*}[h]
\caption {Classified publications concerning the \textit{adversarial} applications of LLMs.}
\label{TableAttack}
\centering
\footnotesize
\resizebox{0.95\textwidth}{!}{
\begin{tabular}{lcllc} 
\hline

\textbf{Paper} &
\textbf{Year} &
\textbf{MITRE Tactic(s)} &
\textbf{Application} &
\textbf{Model(s)} \\
\hline

\cite{charan2023text}   &
2023 &
Execution &
Generating code to perform actions that could be malicious&
GPT-3 \\ 
\hline

\cite{karanjai2022targeted}   &
2022 &
Initial Access   &
Generate phishing emails to bypass spam filters&
GPT-2, GPT-3, RoBERTa  \\ 
\hline

\cite{beckerich2023ratgpt}   &
2022 &
Execution - Command \& Control &
Use of LLMs as plug-ins to act as a proxy &
GPT-4  \\ 
\hline

\cite{saha2023generating}   &
2023 &
Initial Access  - Collection &
Generate Phishing Website via ChatGBT &
GPT-3.5 Turbo \\ 
\hline

\cite{botacin2023gpthreats}   &
2023 &
Execution &
Code generation and DLL injection &
GPT-3 \\ 
\hline

\cite{hazell2023large}   &
2023 &
Initial Access  - Reconnaissance &
Collecting victim data to develop an attack email &
GPT-3.5, GPT-4 \\ 
\hline

\cite{pa2023attacker}   &
2023 &
Initial Access  - Execution - Defense Evasion&
Crafting malicious scripts &
GPT-3.5 Turbo, GPT-4, text-davinci-003 \\ 
\hline

\cite{john1802generative}   &
2018 &
Initial Access &
Spear Phishing link &
AWD-LSTM \\  
\hline

\cite{chatzoglou2023bypassing}   &
2023 &
Defense Evasion &
Code obfuscation, file format modification &
GPT-3.5 \\
\hline

\cite{rando2023passgpt}   &
2023 &
Initial Access  - Credential Access &
Password guessing using LLMs &
GPT-2 \\ 
\hline

\cite{salewski2023context}   &
2023 &
Initial Access  - Reconnaissance &
Impersonation for phishing aims &
GPT-3.5 Turbo \\ 
\hline

\cite{kreps2022all}   &
2022 &
Initial Access &
Generating content for misinformation  &
GPT-2 \\ 
\hline

\end{tabular}

} 
\end{table*}
\subsection{Defense Evasion}
The concept of defense evasion outlines the tactics attackers employ in order to prevent detection following a security breach \cite{xiong2022cyber}. According to Chatzoglou et al. \cite{chatzoglou2023bypassing}, LLMs develop turnkey malware which lets adversaries evade antivirus and endpoint detection and response systems aiming to autonomous malicious code development. Process injection, multiprocessing, junk data, shellcode mem loading, encryption, and chosen shell code were among the techniques employed in their investigation. According to Chatzoglou et al. \cite{chatzoglou2023bypassing} LLMs establish an initial TCP listener that resembles an SSH listener. This will let an attacker to connect and use Windows native APIs to execute Command Prompt (cmd) instructions. An open firewall port is required for the listener to function properly. Only three of the twelve antivirus applications were able to identify malware, according to the author's findings \cite{chatzoglou2023bypassing}. \par
The study conducted by Pa Pa et al. \cite{pa2023attacker} assesses the effectiveness of malware scanners in detecting both obfuscated and non-obfuscated forms of code generated by LLMs. In contrast to LLM-based commonly used obfuscation techniques including base64 encoding or variable and function name  modification, the authors \cite{pa2023attacker} demonstrated that generated non-obfuscated malware featured a reduced detection rate.\par
The use of evasive approaches by LLMs to evade detection by anti-phishing organizations is highlighted by Roy et al. \cite{roy2023generating}. This study illustrates how LLMs assist attackers via clickjacking, fingerprinting browsers, or encoding content. Accordingly, the content of the phishing website is masked using these tactics, making it more challenging for automated anti-phishing crawlers to identify malicious information. \par

\subsection{Credential Access}
Approaches to get credentials through key-logging or credential dumping from a compromised machine refer to credential access \cite{xiong2022cyber}. Introduced by Rando et al. \cite{rando2023passgpt}, PassGPT is an LLM-based password modeling solution. PassGPT uses GPT-2 architecture to estimate password strength and guess passwords. Additionally, the authors \cite{rando2023passgpt} analyze the probability distribution through passwords defined by PassGPT. In light of this, PassGPT delivers guided password generation, enabling constraints to choose character level randomization for the search space by setting parameters like password length or fixed characters with complete control over each character. \par

\subsection{Collection}
Collection refers to gathering information related to the attackers goals \cite{xiong2022cyber}.  Methodologies that demonstrate how LLMs assist in gathering user data are covered by Roy et al. \cite{roy2023generating}. The authors \cite{roy2023generating} investigate the applicability of LLMs in the design of credential taking objects with generating input forms. Furthermore, LLMs have the capability to distribute iFrame injection code to launch malicious websites within an official page. Roy et al. \cite{roy2023generating} demonstrate a scam attack implemented via ChatGPT to gather information without direct attempt aimed at automated data collection. The presented scam item has a hidden iFrame associated with a malicious as well as fake Amazon webpage, guaranteeing that the iFrame object does not activate any anti cross site scripting. \par

\subsection{Command and Control}
Attacks known as command and control arise when an attacker uses a victim channel to connect with underlying resources \cite{xiong2022cyber}.
By leveraging LLMs for performing shell commands on a victim's resource, Beckerich et al. \cite{beckerich2023ratgpt} demonstrate the notion of a command and control attack. In order to generate the executable and automate connection between the machine used by the victim and servers, the authors utilized an LLM-based plugin that acts as an interface for communicating with GPT-2. This method involves utilizing a connectivity feature to establish a connection to a certain website that hosts an attacker's command, followed by a query that ends in a URL. A list of valid user agents used by plugins is maintained regularly in order to mask the malicious component of the web server. \par

Figure \ref{fig3} depicts the study trends on the use of LLMs in cyberattacks, and Table \ref{TableAttack} provides a summary of the categorization. Figure \ref{fig3} illustrates that initial access, defense evasion, and execution tactics are the primary points of concentration for the majority of attack methodologies. As a result, cybersecurity professionals must to give priority to these crucial phases while developing strategic protection methods against LLM-based attacks.

%% file: content/04_future-work-and-conclusion.tex
\section{Conclusion }
\label{conclusion}
In this paper, we reviewed the state-of-the-art research in the applications of Large Language Models (LLMs) within the realm of cybersecurity.
We demonstrated that while LLMs can provide effective solutions for strengthening defensive approaches, their potential misuse cannot be underestimated.
Hence, we categorized related literature using the NIST cybersecurity framework and MITRE attack for applications of LLMs in cyberdefense and cyberattacks, respectively.
Our review suggests that while there are numerous works evaluating the opportunities in defensive applications of LLMs, there is a lack of research in examining the risks of offensive applications.
We hope this study paves the way for future research to assess the associated risks introduced by the rise of LLMs in cybersecurity.